\documentclass[conference]{IEEEtran}
\IEEEoverridecommandlockouts

\usepackage{cite}
\usepackage{amsmath,amssymb,amsfonts}
\usepackage{graphicx}
\usepackage{textcomp}

\usepackage{booktabs}
\usepackage{multirow}
\usepackage{makecell}
\usepackage{array}
\usepackage{tabularx}
\usepackage{colortbl}
\usepackage[table,dvipsnames]{xcolor}
\usepackage{pifont}

\definecolor{BestCell}{RGB}{255,232,232}
\definecolor{SecondCell}{RGB}{232,246,239}
\definecolor{GroupGray}{RGB}{246,246,246}
\definecolor{AblLightGray}{RGB}{246,246,246}
\definecolor{AblBestRed}{RGB}{255,232,232}

\newcommand{\best}[1]{\cellcolor{BestCell}\textbf{#1}}
\newcommand{\second}[1]{\cellcolor{SecondCell}\underline{#1}}
\newcommand{\NA}{--}
\newcommand{\cmark}{\ding{51}}
\newcommand{\dash}{--}
\newcommand{\abest}[1]{\cellcolor{AblBestRed}\textbf{#1}}
\newcommand{\metric}[1]{\footnotesize\textbf{#1}}

\newcolumntype{Y}{>{\centering\arraybackslash}X}
\newcolumntype{M}{>{\raggedright\arraybackslash}p{0.155\textwidth}}

\title{EID-Net: Endoscopic Illumination-aware Degradation Correction Network
}

\author{
\IEEEauthorblockN{
Junzhi Liu\textsuperscript{1,\textdagger},
Yinghao Zeng\textsuperscript{2,\textdagger},
Yang Liu\textsuperscript{1,*},
Huafeng Li\textsuperscript{3},
Xiaosong Li\textsuperscript{1,*}
}
\IEEEauthorblockA{
\textsuperscript{1}\textit{College of Physics and Optoelectronic Engineering, Foshan University,}\\
\textit{Foshan 528225, China}\\
\textsuperscript{2}\textit{School of Computer Science and Engineering, Huizhou University, Huizhou 516007, China}\\
\textsuperscript{3}\,\textit{School of Information Engineering and Automation,}\\
\textit{Kunming University of Science and Technology, Kunming 650500, China}\\
}
\thanks{\textsuperscript{\textdagger}Junzhi Liu and Yinghao Zeng are co-first authors.}%
\thanks{\textsuperscript{*}Corresponding authors.}
}

\begin{document}
\maketitle

\begin{abstract}
Endoscopic images captured under complex illumination conditions often suffer from spatially non-uniform exposure degradation, including local under-exposure, over-exposure, and illumination-induced color variation. 
These degradations are usually entangled with tissue structures, making it challenging for general-purpose low-light enhancement and image restoration methods to improve visibility while preserving local structural and chromatic cues. 
In this paper, we propose EID-Net, an endoscopic illumination-aware degradation correction network for structure-preserving endoscopic illumination correction. 
Rather than treating illumination correction as a global intensity adjustment, EID-Net explicitly extracts an endoscopy-oriented degradation prior from the input image to characterize illumination-related variations in a unified representation. 
This prior is then introduced into the decoder through an illumination-modulated transposed attention mechanism, where it adaptively guides feature aggregation by modulating the key-value feature subspace. 
Extensive experiments on CEC, Endo4IE, EndoVis17, and EndoVis18 datasets demonstrate that EID-Net performs favorably against recent endoscopic enhancement and general image restoration methods. 
The code is available at https://github.com/JUNZHIlius/EID-Net.
\end{abstract}


\section{Introduction}
Endoscopic imaging is an essential tool for gastrointestinal diagnosis and minimally invasive surgery. 
Wireless capsule endoscopy (WCE), as a representative gastrointestinal imaging modality, enables continuous visual inspection of the digestive tract using a swallowable miniaturized imaging device and provides a less invasive way to observe regions that are difficult to fully access with conventional endoscopy. 
However, constrained anatomical cavities, limited onboard imaging hardware, built-in light sources, short imaging distances, and complex tissue geometry often lead to spatially non-uniform illumination degradation in endoscopic images. 
Such degradation may appear as exposure imbalance, color distortion, and reflective artifacts, which can obscure clinically meaningful visual cues and affect downstream computer-aided analysis, such as lesion detection~\cite{xie2022development} and region segmentation~\cite{bai2022discrepancy}.

Endoscopy-assisted minimally invasive surgery also requires stable and clear visual feedback for observing tissue structures, instruments, and operative regions. 
In surgical endoscopic scenes, inadequate illumination and low local contrast can reduce the visibility of anatomical details and make visual guidance more challenging. 
Therefore, robust endoscopic illumination correction is important for improving image visibility while preserving local tissue structures, chromatic consistency, and clinically meaningful visual details.

\begin{figure}[!t]
    \centering
    \includegraphics[width=0.85\linewidth]{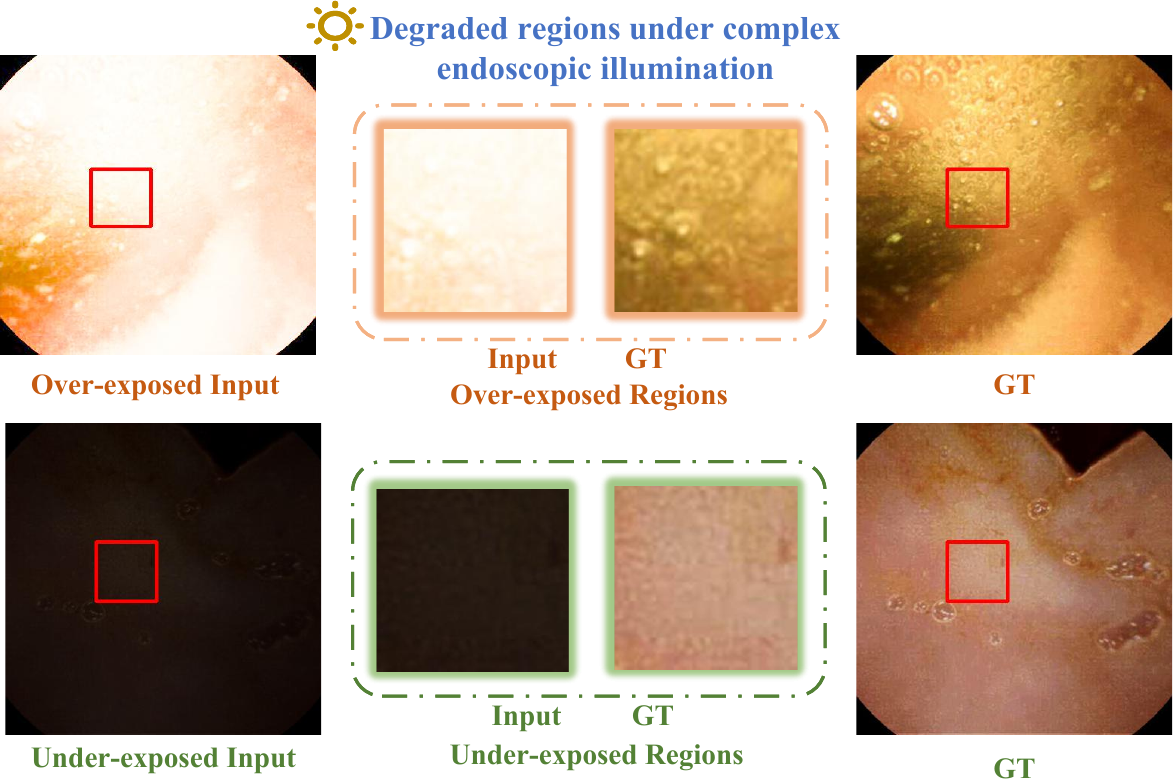}
    \caption{
    Endoscopic images often exhibit spatially localized over-exposure and under-exposure. 
    }
    \label{fig:motivation}
    \vspace{-2mm}
\end{figure}
As shown in Fig.~\ref{fig:motivation}, endoscopic illumination degradation involves more than a
global intensity change. Overexposure can saturate local textures and wash out colors, whereas
underexposure can obscure mucosal structures and vascular details. Global intensity adjustment alone
may leave locally degraded regions insufficiently restored or over-enhance well-exposed areas. These
observations motivate a restoration strategy that uses local illumination cues to guide feature
reconstruction.
Learning-based methods have advanced endoscopic image enhancement and illumination correction. 
More recent methods use diffusion models, prompt learning,
frequency decomposition, or expert routing to handle exposure and illumination
variations~\cite{bai2023llcaps,bai2024endouic,chen2024lightdiff,wu2025prior,ding2026bit}. Despite
these advances, correcting local illumination degradation while preserving tissue structure and
color remains challenging.

To address this problem, we propose EID-Net, an endoscopic illumination-aware degradation correction network for structure-preserving illumination correction. 
Instead of treating illumination correction as a global intensity adjustment, EID-Net formulates endoscopic enhancement as a prior-guided degradation correction problem. 
Specifically, we design an endoscopy-oriented illumination prior estimator to extract luminance distribution, local illumination variation, tissue-related chromatic cues, and specular highlight responses from the degraded input. 
The estimated prior is then injected into the decoder through an illumination-modulated transposed attention mechanism, where it conditionally modulates the key-value feature subspace to guide feature aggregation. 
In this way, EID-Net adapts feature reconstruction to spatially varying illumination degradation while maintaining structural and chromatic fidelity. The main contributions of this work are summarized as follows:

\begin{itemize}
     \item We propose EID-Net, a structure-preserving endoscopic illumination correction network that formulates enhancement as prior-guided degradation correction rather than global brightness adjustment.


     \item We introduce Illumination-Modulated Transposed Attention (IMTA) to condition decoder feature aggregation on the estimated prior through shared key-value modulation.

    \item Extensive experiments on CEC, Endo4IE, EndoVis17, and EndoVis18 demonstrate the effectiveness of EID-Net, while efficiency comparisons show that EID-Net maintains a lightweight computational profile.
    
\end{itemize}

\begin{figure*}[t]
    \centering
    \includegraphics[width=0.95\textwidth]{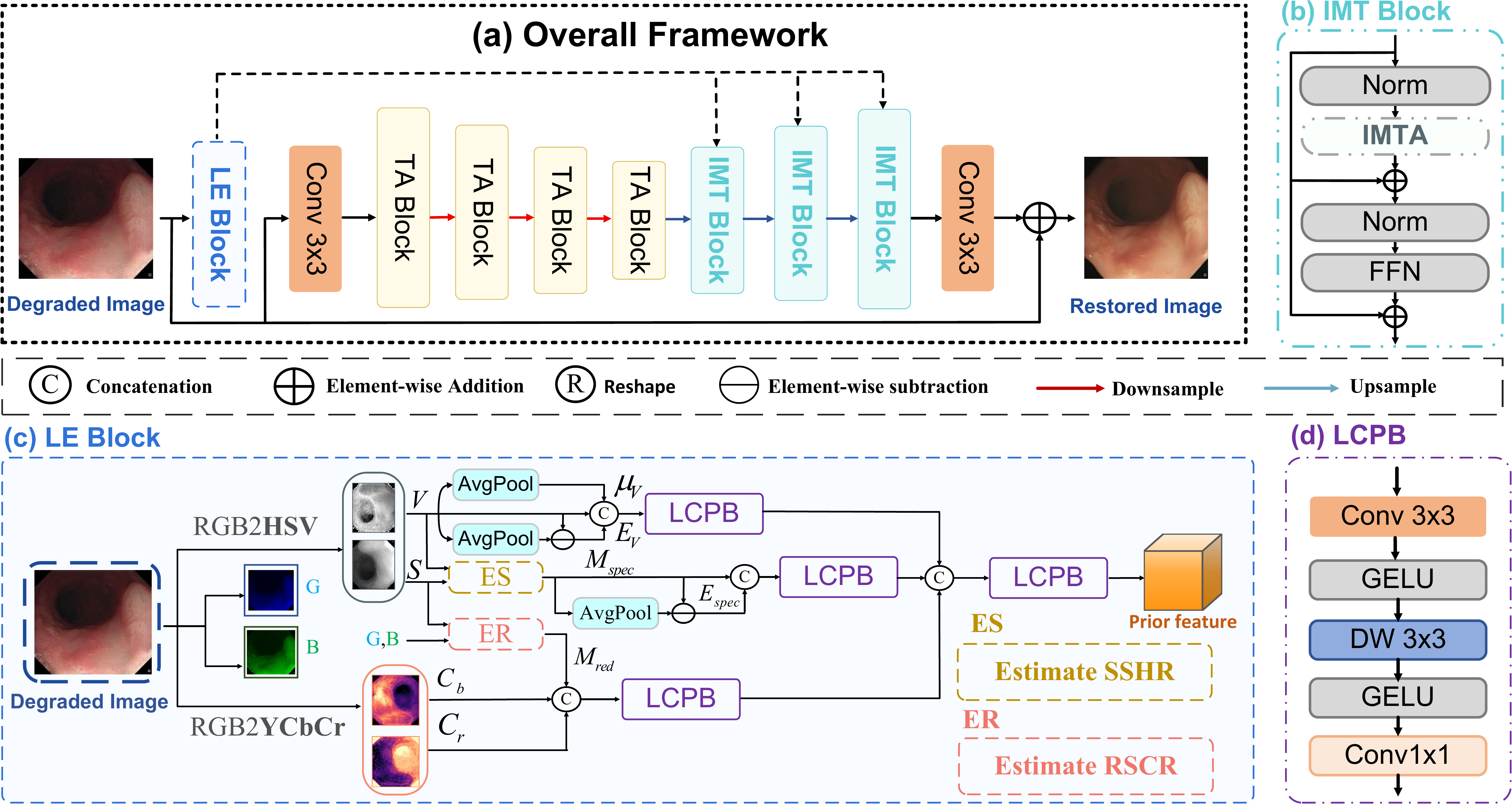}
\caption{
Overview of EID-Net. The U-shaped encoder--decoder combines an illumination prior estimator (LE
Block), transposed attention blocks (TA Blocks)~\cite{zamir2022restormer}, and
illumination-modulated Transformer blocks (IMT Blocks). The LE Block uses Lightweight Convolutional
Projection Blocks (LCPBs) for feature projection. SSHR and RSCR denote the soft specular highlight
response and red-sensitive chromatic response, respectively. Some skip connections are omitted for
clarity.
}
    \label{fig:framework}
\end{figure*}

\section{Related Work}

\subsection{Endoscopic Image Enhancement and Illumination Correction}

Deep learning-based methods have been widely studied for endoscopic image enhancement and illumination correction. 
For low-light WCE enhancement, LLCaps~\cite{bai2023llcaps} combines multi-scale convolution, Curved Wavelet Attention, and a reverse diffusion process to restore local details under insufficient illumination. 
LighTDiff~\cite{chen2024lightdiff} further explores diffusion-based enhancement for surgical endoscopic images, to improve restoration efficiency and reduce chromatic bias.
Recent studies have also considered more complex illumination states in endoscopic imaging. 
EndoUIC~\cite{bai2024endouic} addresses unified illumination correction for WCE images by introducing illumination prompts into a diffusion Transformer, enabling adaptive correction of under-exposed and over-exposed images. 
WTNet~\cite{wu2025prior} exploits the observation that illumination information is mainly represented in low-frequency components, and builds a lightweight exposure correction network based on wavelet-domain priors. 
BiT-MoFE~\cite{ding2026bit} combines Brownian Bridge Diffusion with frequency-based expert routing to recover illumination and structural information in a more adaptive manner. 
Beyond illumination degradation, endoscopic restoration may also involve scene-specific artifacts such as specular highlights, smoke, and bleeding. 
FAQNet~\cite{zhu2024faqnet} focuses on endoscopic highlight removal caused by built-in light sources and mucosal reflection, while EndoIR~\cite{chen2026endoir} formulates endoscopic image restoration as an all-in-one problem for multiple degradations.
These methods show that endoscopic enhancement has gradually moved from simple contrast or brightness adjustment toward degradation-aware restoration. 

\subsection{Guided Attention Modeling and Feature Modulation}

Transformer-based architectures have been widely used in image restoration~\cite{zamir2022restormer,liang2021swinir}.
Instead of relying only on fixed feature mappings, recent restoration methods introduce priors, prompts, or modulation mechanisms to make the restoration process adaptive to the degradation condition. 
DynRouteNet~\cite{zeng2026dynroutenet} uses shadow masks to
condition dynamic routing for facial shadow removal, whereas
AquaClarity~\cite{chen2026aquaclarity} combines channel and
spatial attention with nonlinear feature fusion and color
refinement for underwater enhancement.
Face Shadow Eraser (FSE)~\cite{11462271} combines hierarchical
shifted-window attention with feature modulation conditioned
on shadow masks to refine local illumination in facial
shadow removal.
These designs highlight the role of task-specific cues and
representations in feature processing.
Retinexformer~\cite{cai2023retinexformer} estimates illumination representations under a Retinex-based framework and uses an illumination-guided Transformer to model interactions among regions with different lighting conditions. 
PromptIR~\cite{potlapalli2023promptir} uses learnable prompts to represent degradation information and guide the network to handle different restoration tasks. 
MoCE-IR~\cite{zamfir2025complexity} introduces expert selection based on degradation-related embeddings, allowing different restoration paths for different degradation types. 
AdaIR~\cite{cui2025adair} further explores adaptive feature modulation by promoting interactions between high- and low-frequency components during image restoration.



\section{Method}

\subsection{Overall Framework}

Given a degraded endoscopic image $\mathbf{I}\in\mathbb{R}^{H\times W\times3}$, EID-Net aims to obtain a restored image $\hat{\mathbf{I}}\in\mathbb{R}^{H\times W\times3}$ with corrected illumination and preserved tissue structures. 
As shown in Fig.~\ref{fig:framework}, EID-Net adopts a U-shaped architecture~\cite{ronneberger2015u}. 
In parallel, an illumination prior estimator, termed LE Block, extracts an endoscopy-oriented illumination prior $\mathbf{P}$ from the degraded input. 
This prior is not directly used to adjust image intensity, but is introduced into the decoder to guide feature reconstruction under spatially varying illumination degradation.
The encoder is composed of transposed attention blocks~\cite{zamir2022restormer}, denoted as TA Blocks, and progressively converts high-resolution shallow features into low-resolution latent representations. 
Pixel-Unshuffle is used for downsampling between adjacent encoder stages. 
The decoder mirrors the encoder structure and reconstructs high-resolution features through Pixel-Shuffle upsampling and skip connections from the corresponding encoder stages. 
Finally, a $3\times3$ convolution maps the reconstructed features back to the image space. 
Following common practice in image restoration, EID-Net predicts a residual image $\Delta\mathbf{I}$ and obtains the final output by $\hat{\mathbf{I}}=\mathbf{I}+\Delta\mathbf{I}$. 

\subsection{Illumination Prior Estimator Block}
The Illumination Prior Estimator Block (LE Block) extracts luminance, chromatic, and specular cues
from the input image. Unless otherwise specified, each projection $\Phi(\cdot)$ uses a LCPB, the composition of which is shown in the Fig~\ref{fig:framework} (d).

Given a degraded RGB image $\mathbf{I}\in[0,1]^{H\times W\times3}$ with color channels $\mathbf{R}$, $\mathbf{G}$, and $\mathbf{B}$, we derive the HSV value and saturation components as $\mathbf{V}=\max(\mathbf{R},\mathbf{G},\mathbf{B})$ and $\mathbf{S}=(\mathbf{V}-\min(\mathbf{R},\mathbf{G},\mathbf{B}))/(\mathbf{V}+\epsilon)$, where $\epsilon$ is a small constant for numerical stability. 
$\mathbf{V}$ serves as the luminance response, while $\mathbf{S}$ provides complementary saturation information.
To capture local illumination distribution and brightness variation, we compute
\begin{equation}
\begin{aligned}
\boldsymbol{\mu}_{v}
&=
\operatorname{AvgPool}_{7\times7}(\mathbf{V}), \\
\mathbf{E}_{v}
&=
\left|
\mathbf{V}
-
\operatorname{AvgPool}_{3\times3}(\mathbf{V})
\right|, \\
\mathbf{F}_{\mathrm{lum}}
&=
\Phi_{\mathrm{lum}}
\left(
[
\mathbf{V},
\boldsymbol{\mu}_{v},
\mathbf{E}_{v}
]
\right).
\end{aligned}
\label{eq:lum_prior}
\end{equation}
Here, $\boldsymbol{\mu}_{v}$ and $\mathbf{E}_{v}$ describe local
brightness distribution and short-range luminance variation,
respectively.
For tissue-related chromatic cues, we use a YCbCr-based
conversion without chroma offsets to obtain the luminance
component $\mathbf{Y}$ and color-difference components
$\mathbf{C}_{b}$ and $\mathbf{C}_{r}$ from the RGB image.
We also compute a red-sensitive chromatic response (RSCR), $\mathbf{M}_{\mathrm{red}}$, to emphasize
red-dominant, saturated regions:
\begin{equation}
\mathbf{M}_{\mathrm{red}}
=
\sigma
\left(
\frac{
\mathbf{R}-\max(\mathbf{G},\mathbf{B})-\tau_{r}
}{
\eta
}
\right)
\odot
\sigma
\left(
\frac{
\mathbf{S}-\tau_{s}
}{
\eta
}
\right),
\label{eq:red_prior}
\end{equation}
where $\sigma(\cdot)$ is the sigmoid function, $\odot$ denotes element-wise multiplication, $\tau_r$ and $\tau_s$ are fixed thresholds, and $\eta$ controls the softness of the response. 
The chromatic prior feature is generated as
\begin{equation}
\mathbf{F}_{\mathrm{chr}}
=
\Phi_{\mathrm{chr}}
\left(
[
\mathbf{C}_{b},
\mathbf{C}_{r},
\mathbf{M}_{\mathrm{red}}
]
\right).
\label{eq:chr_prior}
\end{equation}

Specular highlights in endoscopic images usually exhibit high brightness and low saturation. 
We therefore compute a soft specular highlight response (SSHR), denoted as $\mathbf{M}_{\mathrm{spec}}$:
\begin{equation}
\mathbf{M}_{\mathrm{spec}}
=
\sigma
\left(
\frac{
\mathbf{V}-\tau_{v}
}{
\eta
}
\right)
\odot
\sigma
\left(
\frac{
\tau_{\mathrm{sat}}-\mathbf{S}
}{
\eta
}
\right),
\label{eq:spec_prior_map}
\end{equation}
where $\tau_v$ and $\tau_{\mathrm{sat}}$ are fixed brightness and saturation thresholds, respectively. 
To capture local transitions around highlight regions, we further compute
\begin{equation}
\begin{aligned}
\mathbf{E}_{\mathrm{spec}}
&=
\left|
\mathbf{M}_{\mathrm{spec}}
-
\operatorname{AvgPool}_{3\times3}
(\mathbf{M}_{\mathrm{spec}})
\right|, \\
\mathbf{F}_{\mathrm{spec}}
&=
\Phi_{\mathrm{spec}}
\left(
[
\mathbf{M}_{\mathrm{spec}},
\mathbf{E}_{\mathrm{spec}}
]
\right).
\end{aligned}
\label{eq:spec_feature}
\end{equation}

The thresholds and softness parameter in the red-sensitive and specular responses are fixed in the normalized RGB space and shared across all datasets.
Finally, the luminance, chromatic, and specular prior features are concatenated and fused to obtain the illumination prior:
\begin{equation}
\mathbf{P}
=
\Phi_{\mathrm{fuse}}
\left(
[
\mathbf{F}_{\mathrm{lum}},
\mathbf{F}_{\mathrm{chr}},
\mathbf{F}_{\mathrm{spec}}
]
\right).
\label{eq:prior_fusion}
\end{equation}
The resulting prior $\mathbf{P}\in\mathbb{R}^{H\times W\times C}$ is later resized and projected to match each decoder stage. 

\begin{figure}[!t]
    \centering
    \includegraphics[
        width=0.60\columnwidth,
        keepaspectratio
    ]{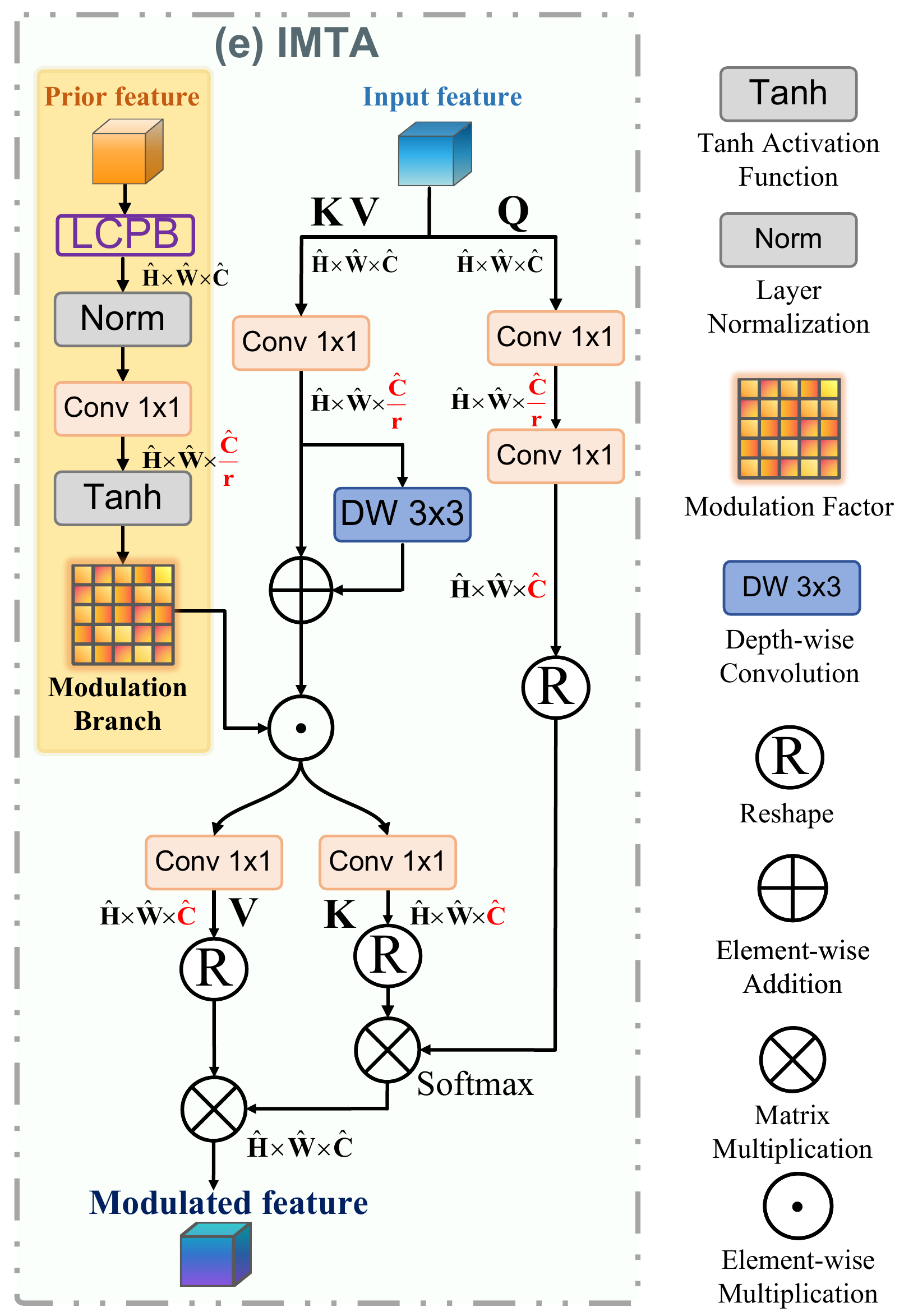}
   
   \caption{
Illustration of the proposed Illumination-Modulated Transposed Attention (IMTA). 
}
    
    \label{fig:imta}
    \vspace{-2mm}
\end{figure}
\subsection{Illumination-Modulated Transformer Block}
The decoder of EID-Net is built with IMT Blocks. 
Each IMT Block follows the standard Transformer-style block design, consisting of normalization layers, an attention module, residual connections, and a gated depth-wise convolutional feed-forward network~\cite{zamir2022restormer}. 
As illustrated in Fig~\ref{fig:framework} (b) and Fig.~\ref{fig:imta}, unlike conventional Transformer blocks that derive attention solely from image features, the key component of the IMT Block is the proposed IMTA, which injects the estimated illumination prior into the attention computation.
Given the normalized decoder feature at the $l$-th stage as 
$\mathbf{X}_{l}\in\mathbb{R}^{H_l\times W_l\times C_l}$ and the illumination prior produced by the LE Block as $\mathbf{P}$, we first resize $\mathbf{P}$ to the spatial resolution of $\mathbf{X}_{l}$ and align it with a lightweight convolutional adapter:
\begin{equation}
\mathbf{P}_{l}
=
\mathcal{A}_{l}
\left(
\operatorname{Resize}
\left(
\mathbf{P}
\right)
\right),
\label{eq:prior_align}
\end{equation}
where $\mathcal{A}_{l}(\cdot)$ denotes the prior adapter at the $l$-th decoder stage, and $\mathbf{P}_{l}$ is the aligned illumination prior.
IMTA contains a query branch and a shared key-value branch. 
The query branch is directly generated from the decoder feature:
\begin{equation}
\mathbf{F}_{q}
=
W_{q}^{\mathrm{down}}
\left(
\mathbf{X}_{l}
\right),
\label{eq:imta_query}
\end{equation}
where $W_{q}^{\mathrm{down}}(\cdot)$ denotes a $1\times1$ convolutional projection. 
For the key-value branch, $\mathbf{X}_{l}$ is first projected into a shared latent subspace, followed by a depth-wise convolution to introduce local context:
\begin{equation}
\begin{aligned}
\mathbf{U}_{kv}
&=
W_{kv}^{\mathrm{down}}
\left(
\mathbf{X}_{l}
\right), \\
\mathbf{F}_{kv}
&=
\mathbf{U}_{kv}
+
\operatorname{DWConv}_{3\times3}
\left(
\mathbf{U}_{kv}
\right),
\end{aligned}
\label{eq:imta_kv}
\end{equation}
where $W_{kv}^{\mathrm{down}}(\cdot)$ is a $1\times1$ convolutional projection. 
The shared key-value projection avoids redundant feature projections, while the depth-wise convolution supplements local spatial context for fine texture reconstruction.
The aligned illumination prior is used to generate a multiplicative modulation factor:
\begin{equation}
\boldsymbol{\Gamma}_{l}
=
1
+
\lambda
\cdot
\tanh
\left(
W_{p}
\left(
\mathbf{P}_{l}
\right)
\right),
\label{eq:imta_gate}
\end{equation}
where $W_p(\cdot)$ maps the aligned prior to the key-value latent space, and $\lambda$ is a learnable scaling factor.
The bounded response of $\tanh(\cdot)$ stabilizes the modulation and allows the key-value features to be enhanced or suppressed according to the estimated illumination prior. 
The modulated key-value feature is then obtained as
\begin{equation}
\widetilde{\mathbf{F}}_{kv}
=
\mathbf{F}_{kv}
\odot
\boldsymbol{\Gamma}_{l},
\label{eq:imta_modulation}
\end{equation}
where $\odot$ denotes element-wise multiplication.
We modulate the key-value branch rather than the query branch because the query should preserve the structural matching request of the current feature, whereas the key and value features determine where to retrieve information and what content to aggregate. 
After modulation, the query, key, and value tensors are generated by
\begin{equation}
\mathbf{Q}=W_q^{u}(\mathbf{F}_q),\quad
\mathbf{K}=W_k^{u}(\widetilde{\mathbf{F}}_{kv}),\quad
\mathbf{V}=W_v^{u}(\widetilde{\mathbf{F}}_{kv})
\label{eq:imta_qkv}
\end{equation}
where $W_q^{u}(\cdot)$, $W_k^{u}(\cdot)$, and $W_v^{u}(\cdot)$ are $1\times1$ up-projection
convolutions.
Following transposed attention~\cite{zamir2022restormer}, we split $\mathbf{Q}$, $\mathbf{K}$, and
$\mathbf{V}$ into $N$ heads and reshape each head for channel-wise attention. For head $j$,
$\mathbf{Q}_{j}$ and $\mathbf{K}_{j}$ are $\ell_2$-normalized to obtain $\widehat{\mathbf{Q}}_{j}$
and $\widehat{\mathbf{K}}_{j}$. Attention is computed in the channel covariance space as
\begin{equation}
\mathbf{A}_{j}
=
\operatorname{Softmax}
\left(
\frac{
\widehat{\mathbf{Q}}_{j}
\widehat{\mathbf{K}}_{j}^{T}
}{
\alpha_{j}
}
\right),
\label{eq:imta_attn}
\end{equation}
where $\alpha_{j}$ is a learnable temperature parameter for the $j$-th head. 
The illumination-modulated attention output is then computed as
\begin{equation}
\widehat{\mathbf{X}}_{l}
=
W_{o}
\left(
\operatorname{Concat}
\left[
\mathbf{A}_{1}\mathbf{V}_{1},
\ldots,
\mathbf{A}_{N}\mathbf{V}_{N}
\right]
\right)
+
\mathbf{X}_{l},
\label{eq:imta_output}
\end{equation}
where $W_{o}(\cdot)$ is the output projection.
The normalization, residual connection, and feed-forward network in the IMT Block follow the standard Restormer design. 
By conditioning the key-value feature space on the illumination prior, IMTA enables the decoder to adapt feature aggregation to local illumination degradation while preserving structural and chromatic information.

\section{Experiments}
\subsection{Implementation Details}

\subsubsection{Datasets}
We evaluated EID-Net on four endoscopic illumination correction and enhancement datasets, including CEC~\cite{bai2024endouic}, Endo4IE~\cite{garcia2022novel}, EndoVis17~\cite{allan20192017}, and EndoVis18~\cite{allan20202018}. 
CEC and Endo4IE were used to evaluate exposure correction under under-exposed and over-exposed conditions. 
The CEC dataset contains paired properly exposed and degraded images, where 800 pairs were used for training and 200 pairs for testing. 
\begin{table*}[!t]
\centering
\caption{
Quantitative comparison on CEC and Endo4IE datasets. 
The best and second-best results are highlighted with red and green shading, respectively. 
$\uparrow$ indicates higher is better, while $\downarrow$ indicates lower is better.
}
\label{tab:sota_comparison}
\vspace{-1.5mm}

\begingroup
\normalsize
\setlength{\tabcolsep}{2.2pt}
\renewcommand{\arraystretch}{0.82}
\setlength{\aboverulesep}{0.35pt}
\setlength{\belowrulesep}{0.35pt}
\setlength{\heavyrulewidth}{0.65pt}
\setlength{\lightrulewidth}{0.40pt}
\setlength{\cmidrulewidth}{0.40pt}

\begin{tabularx}{0.98\textwidth}{@{}MYYY|YYY@{}}
\toprule
\multirow{2}{*}{\textbf{Method}} 
& \multicolumn{3}{c|}{\textbf{CEC}} 
& \multicolumn{3}{c}{\textbf{Endo4IE}} \\
\cmidrule(lr){2-4} \cmidrule(l){5-7}
\addlinespace[1.2pt]
& \rule{0pt}{2.2ex}\textbf{PSNR$\uparrow$} 
& \rule{0pt}{2.2ex}\textbf{SSIM$\uparrow$} 
& \rule{0pt}{2.2ex}\textbf{LPIPS$\downarrow$} 
& \rule{0pt}{2.2ex}\textbf{PSNR$\uparrow$} 
& \rule{0pt}{2.2ex}\textbf{SSIM$\uparrow$} 
& \rule{0pt}{2.2ex}\textbf{LPIPS$\downarrow$} \\
\midrule

\rowcolor{GroupGray}
\multicolumn{7}{c}{\textit{General image restoration and enhancement methods}} \\
\midrule

MIRNetv2~\cite{zamir2022learning}      
& 28.36 & 93.58 & 0.1080 
& 23.85 & 82.33 & 0.2376 \\

PromptIR~\cite{potlapalli2023promptir}      
& 28.27 & 83.14 & 0.0717 
& 23.73 & 79.57 & 0.2396 \\

AdaIR~\cite{cui2025adair}         
& 34.48 & 96.92 & 0.0413    
& 26.33 & 83.91 & 0.2107 \\

Retinexformer~\cite{cai2023retinexformer}
& \second{37.87} & \second{97.54} & 0.0328
& 28.00 & 85.13 & 0.1939 \\

MoCE-IR~\cite{zamfir2025complexity}       
& 33.33 & 95.51 & 0.1031 
& 27.57 & 84.66 & 0.2086 \\


PyDiff~\cite{zhou2023pyramid}        
& 28.18 & 95.76 & 0.0941 
& 24.73 & 84.78 & 0.2148 \\

LACT~\cite{baek2023luminance}          
& 28.40 & 93.09 & 0.1103 
& 22.92 & 76.88 & 0.2671 \\

LA-Net~\cite{yang2023learning}        
& \NA & \NA & \NA    
& 23.51 & 83.78 & \best{0.1186} \\

\midrule
\rowcolor{GroupGray}
\multicolumn{7}{c}{\textit{Endoscopic image enhancement and illumination correction methods}} \\
\midrule

LLCaps~\cite{bai2023llcaps}        
& 27.55 & 85.95 & 0.2366 
& 22.85 & 66.42 & 0.3446 \\

EndoUIC~\cite{bai2024endouic}       
& 29.65 & 96.80 & 0.0655 
& 25.49 & 85.20 & 0.1937 \\

WTNet~\cite{wu2025prior}         
& 36.82 & 97.43 & 0.0343 
& 27.91 & \second{86.14} & 0.1954 \\

LightDiff~\cite{chen2024lightdiff}     
& 31.95 & 95.23 & 0.0650 
& \best{29.22} & 85.29 & 0.1731 \\

EndoIR~\cite{chen2026endoir}        
& 31.28 & 96.84 & 0.0442 
& 25.37 & 83.59 & 0.1767 \\

BiT-MoFE~\cite{ding2026bit}      
& 35.31 & 97.47 & \second{0.0299} 
& 28.11 & 85.61 & \second{0.1589} \\


\midrule
\textbf{EID-Net (ours)} 
& \best{39.05} & \best{98.33} & \best{0.0249} 
& \second{28.71} & \best{87.34} & 0.1781 \\

\bottomrule
\end{tabularx}


\endgroup
\vspace{-2mm}
\end{table*}
The Endo4IE dataset contains 956 under-exposed and 1,194 over-exposed images, with 1,552 paired images for training and 598 paired images for testing. 
Following LighTDiff~\cite{chen2024lightdiff}, we also used the synthesized low-light versions of the public surgical endoscopic datasets EndoVis17 and EndoVis18 to evaluate low-light enhancement performance in surgical endoscopic scenes.

\subsubsection{Experimental Setup}
EID-Net was implemented in PyTorch and trained end-to-end. 
The shallow feature dimension was set to 32, and the numbers of blocks and attention heads across the four encoder-decoder stages were set to $[1,1,1,1]$ and $[1,2,4,8]$, respectively. 
During training, paired images were randomly cropped into $128\times128$ patches and augmented with random flips and rotations.
We used the Adam optimizer with $\beta_1=0.9$ and $\beta_2=0.99$. 
The initial learning rate was set to $2\times10^{-4}$ and gradually decreased to $1\times10^{-6}$. 
The model was trained for 1,500 epochs with a batch size of 8. 
We optimized the network using a weighted combination of the $\ell_1$ reconstruction loss and the SSIM loss, defined as $\mathcal{L}=\mathcal{L}_{1}+0.2\mathcal{L}_{\mathrm{SSIM}}$.

\begin{figure*}[!t]
    \centering
    \includegraphics[
        width=0.96\textwidth
    ]{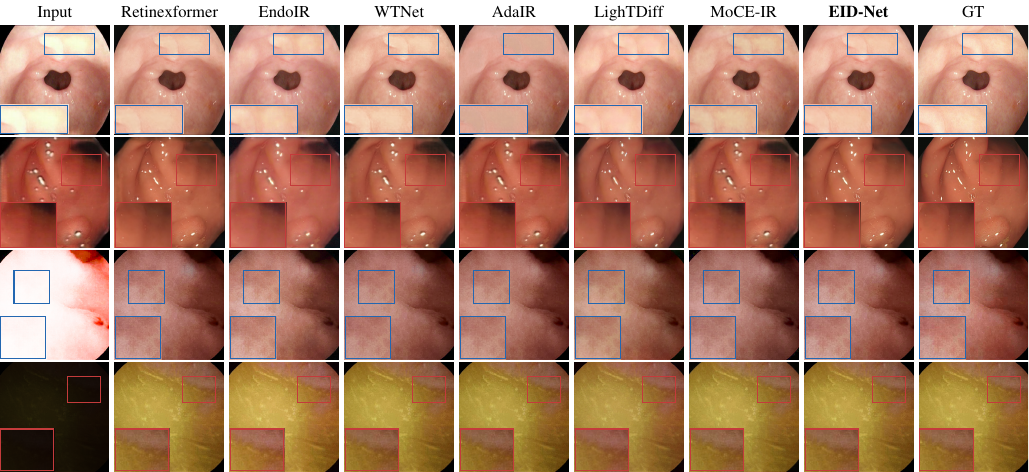}
    \caption{
    Visual comparison on Endo4IE and CEC.
    From top to bottom: overexposed and underexposed
    Endo4IE samples, followed by overexposed and
    underexposed CEC samples.
    }
    \label{fig:visual_comparison_endo4ie_cec}
\end{figure*}

\begin{figure*}[!t]
    \centering
    \includegraphics[
        width=0.96\textwidth
    ]{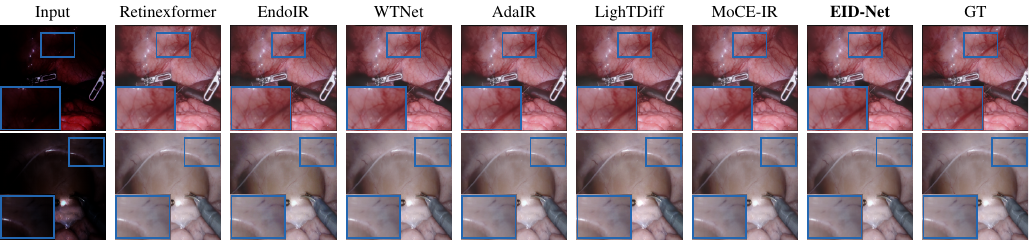}
    \caption{
    Visual comparison on EndoVis17 (top)
    and EndoVis18 (bottom).
    }
    \label{fig:endovis_visual_comparison}
\end{figure*}




\subsection{Comparison with State-of-the-Art Methods}

\subsubsection{Quantitative Comparison}

To comprehensively evaluate the effectiveness of EID-Net, we compare it with 14 representative competing methods.
All methods are evaluated on the same test sets using three full-reference metrics: Peak Signal-to-Noise Ratio (PSNR),
Structural Similarity (SSIM),
and Learned Perceptual Image Patch Similarity (LPIPS).
The quantitative results on CEC and Endo4IE are reported in Table~\ref{tab:sota_comparison}.
On CEC, EID-Net achieves the best performance across all three metrics, obtaining 39.05 dB PSNR, 98.33 SSIM, and 0.0249 LPIPS. 
Compared with Retinexformer and WTNet, which also achieve strong results on this dataset, EID-Net provides higher pixel fidelity and better perceptual quality, indicating its effectiveness in correcting non-uniform exposure degradation. 
On Endo4IE, EID-Net achieves the best SSIM and the second-best PSNR. 
Although LA-Net obtains the lowest LPIPS and LighTDiff achieves the highest PSNR on this dataset, EID-Net shows a better balance between structural similarity and exposure correction, demonstrating robust performance under both under-exposed and over-exposed conditions.
Table~\ref{tab:endovis_comparison} reports the results on EndoVis17 and EndoVis18. 
On EndoVis17, EID-Net achieves the best PSNR, SSIM, and LPIPS, showing its ability to enhance low-light surgical endoscopic images while preserving local structures. 
On EndoVis18, EID-Net obtains the best PSNR and LPIPS, and achieves the second-best SSIM, slightly lower than WTNet. 
These results suggest that the proposed illumination prior and IMTA mechanism generalize well from exposure correction datasets to low-light surgical endoscopic enhancement scenarios.

\begin{table}[!t]
\centering
\caption{
Quantitative comparison on EndoVis17 and EndoVis18. 
Best and second-best results are shaded in red and green.
}
\label{tab:endovis_comparison}
\vspace{-1mm}

\begingroup
\scriptsize
\setlength{\tabcolsep}{2.2pt}
\renewcommand{\arraystretch}{0.94}
\setlength{\aboverulesep}{0.35pt}
\setlength{\belowrulesep}{0.35pt}
\setlength{\heavyrulewidth}{0.65pt}
\setlength{\lightrulewidth}{0.42pt}
\setlength{\cmidrulewidth}{0.42pt}

\resizebox{0.98\columnwidth}{!}{%
\begin{tabular}{@{}lccc|cc>{\columncolor{white}[\tabcolsep][0pt]}c@{}}
\toprule
\multirow{2}{*}{\textbf{Method}}
& \multicolumn{3}{c|}{\textbf{EndoVis17}}
& \multicolumn{3}{c}{\textbf{EndoVis18}} \\
\cmidrule(lr){2-4} \cmidrule(l){5-7}
& \textbf{PSNR$\uparrow$}
& \textbf{SSIM$\uparrow$}
& \textbf{LPIPS$\downarrow$}
& \textbf{PSNR$\uparrow$}
& \textbf{SSIM$\uparrow$}
& \textbf{LPIPS$\downarrow$} \\
\midrule

AdaIR
& \second{36.10} & 96.48 & \second{0.0274}
& \second{33.97} & 95.83 & \second{0.0399} \\

Retinexformer
& 33.89 & 95.91 & 0.0359
& 33.23 & 95.56 & 0.0451 \\

MIRNetv2
& 31.94 & 93.48 & 0.0671
& 28.96 & 92.63 & 0.0799 \\

LLCaps
& 31.88 & 93.53 & 0.0589
& 25.18 & 90.16 & 0.1110 \\

WTNet
& 34.36 & \second{96.70} & 0.0328
& 33.48 & \best{96.28} & 0.0437 \\

EndoIR
& 32.49 & 95.62 & 0.0346
& 29.86 & 94.67 & 0.0504 \\

MoCE-IR
& 33.81 & 96.56 & 0.1047
& 32.75 & 96.02 & 0.1196 \\

PyDiff
& 31.02 & 94.50 & 0.0510
& 29.37 & 93.82 & 0.0711 \\

LighTDiff
& 34.28 & 95.72 & 0.0325
& 31.99 & 94.91 & 0.0531 \\

\midrule
\textbf{EID-Net}
& \best{36.13} & \best{96.74} & \best{0.0253}
& \best{34.04} & \second{96.06} & \best{0.0363} \\

\bottomrule
\end{tabular}}

\endgroup
\vspace{-2mm}
\end{table}

\subsubsection{Qualitative Comparison}

Fig.~\ref{fig:visual_comparison_endo4ie_cec} shows visual comparisons on the CEC and Endo4IE datasets. We notice that
the compared methods either leave locally under-exposed regions insufficiently restored or introduce over-enhancement in bright regions. 
In contrast, EID-Net improves local visibility while maintaining natural global brightness and tissue color consistency. 
Fig.~\ref{fig:endovis_visual_comparison} shows visual comparisons on EndoVis17 and EndoVis18. 
Compared with existing methods, EID-Net restores clearer tissue structures and vascular details while reducing low-light degradation. 
It also maintains more stable chromatic appearance in surgical endoscopic scenes, avoiding obvious color shifts or excessive brightness amplification. 
These qualitative results are consistent with the quantitative improvements reported in Table~\ref{tab:endovis_comparison}.

\subsection{Ablation Study and Model Complexity}

We conducted ablation studies on Endo4IE to evaluate the proposed illumination-modulated attention target and illumination prior components, as shown in Table~\ref{tab:ablation}.
The baseline removes the proposed illumination-guided modulation while keeping the same Restormer-style backbone~\cite{zamir2022restormer}. 
To verify the modulation target in IMTA, we separately modulate the query, key, value, and key-value branches. 
Guided-KV achieves the best PSNR and SSIM while maintaining competitive LPIPS, showing a better balance between fidelity and perceptual quality. 
The full prior with all three cues achieves the best overall performance, demonstrating the effectiveness of the proposed illumination-aware prior for endoscopic degradation correction. Table~\ref{tab:complexity} compares model complexity and inference efficiency with representative methods. EID-Net achieves the lowest FLOPs and the highest FPS among them. 

\begin{table}[!t]
\centering
\caption{
Ablation study on the attention modulation target and illumination prior components. 
}
\label{tab:ablation}
\vspace{-1mm}

\begingroup
\normalsize
\setlength{\tabcolsep}{2.3pt}
\renewcommand{\arraystretch}{0.90}
\setlength{\aboverulesep}{0.35pt}
\setlength{\belowrulesep}{0.35pt}
\setlength{\heavyrulewidth}{0.65pt}
\setlength{\lightrulewidth}{0.42pt}

\begin{tabular}{@{}lccc ccc@{}}
\toprule
\textbf{Setting} 
& \metric{Q} 
& \metric{K} 
& \metric{V} 
& \metric{PSNR$\uparrow$} 
& \metric{SSIM$\uparrow$} 
& \metric{LPIPS$\downarrow$} \\
\midrule
\rowcolor{AblLightGray}
\multicolumn{7}{c}{\textit{Effect of illumination-modulated attention target}} \\
\midrule

Baseline  
& \dash & \dash & \dash 
& 26.99 & 85.06 & 0.2093 \\

Guided-V  
& \dash & \dash & \cmark 
& 28.70 & 87.26 & \abest{0.1772} \\

Guided-Q  
& \cmark & \dash & \dash 
& 28.64 & 87.19 & 0.1782 \\

Guided-K  
& \dash & \cmark & \dash 
& 28.65 & 87.16 & 0.1787 \\

Guided-KV 
& \dash & \cmark & \cmark 
& \abest{28.71} & \abest{87.34} & 0.1781 \\

\midrule
\textbf{Setting} 
& \metric{Light} 
& \metric{Color} 
& \metric{Spec.} 
& \metric{PSNR$\uparrow$} 
& \metric{SSIM$\uparrow$} 
& \metric{LPIPS$\downarrow$} \\
\midrule
\rowcolor{AblLightGray}
\multicolumn{7}{c}{\textit{Effect of illumination prior components}} \\
\midrule

w/o Prior 
& \dash & \dash & \dash 
& 28.67 & 87.19 & 0.1785 \\

Light only 
& \cmark & \dash & \dash 
& 28.64 & 87.13 & 0.1785 \\

Light + Color 
& \cmark & \cmark & \dash 
& 28.65 & 87.21 & 0.1794 \\

Light + Color + Spec. 
& \cmark & \cmark & \cmark 
& \abest{28.71} & \abest{87.34} & \abest{0.1781} \\


\bottomrule
\end{tabular}

\endgroup
\vspace{-2mm}
\end{table}

\begin{table}[!t]
\centering
\caption{
Efficiency comparison with recent endoscopic restoration methods. 
FLOPs and FPS are measured at $128\times128$ resolution.
}
\label{tab:complexity}
\vspace{-1mm}

\begingroup
\normalsize
\setlength{\tabcolsep}{3.0pt}
\renewcommand{\arraystretch}{0.95}
\setlength{\aboverulesep}{0.35pt}
\setlength{\belowrulesep}{0.35pt}
\setlength{\heavyrulewidth}{0.65pt}
\setlength{\lightrulewidth}{0.42pt}

\begin{tabular*}{0.98\columnwidth}{@{\extracolsep{\fill}}lccc@{}}
\toprule
\textbf{Method} 
& \textbf{Params (M)$\downarrow$} 
& \textbf{FLOPs (G)$\downarrow$} 
& \textbf{FPS$\uparrow$} \\
\midrule

WTNet~\cite{wu2025prior}     
& \textbf{1.40} 
& \underline{11.28} 
& \underline{21.2} \\

EndoIR~\cite{chen2026endoir}    
& 20.30 
& 30.61 
& 10.6 \\

LightDiff~\cite{chen2024lightdiff} 
& 18.78 
& 41.93 
& 21.1 \\

\midrule
\textbf{EID-Net} 
& \underline{3.62} 
& \textbf{4.47} 
& \textbf{80.5} \\

\bottomrule
\end{tabular*}

\endgroup
\vspace{-2mm}
\end{table}












\section{Conclusion}
In this paper, we proposed EID-Net, an illumination-aware degradation correction network for structure-preserving endoscopic image enhancement. 
EID-Net extracts an endoscopy-oriented degradation prior from luminance, chromatic, and specular cues, and incorporates it into the decoder through illumination-modulated transposed attention to adapt feature aggregation to spatially varying illumination degradation. 
Extensive experiments 
demonstrate that EID-Net achieves competitive performance against recent endoscopic enhancement and general image restoration methods while maintaining a lightweight computational profile. 
Ablation studies further validate the effectiveness of the proposed key-value modulation and estimated prior for robust illumination correction. 
Future work will investigate the robustness of EID-Net under more diverse real-world endoscopic acquisition conditions.

\section{Acknowledgement}
This work was supported by the Guangdong Basic and Applied Basic Research Foundation (Nos. 2023A1515140077 and 2024A1515011880), and the Yunnan Fundamental Research Projects (Nos. 202301AV070004 and 202501AS070123) .

\bibliographystyle{IEEEtran}
\bibliography{bibm_ref}

\end{document}